\documentclass[iop]{emulateapj}
\usepackage{times,color,natbib,url,amsmath}
\usepackage[colorlinks,urlcolor=blue,citecolor=blue,linkcolor=blue]{hyperref} 

\newenvironment{ant-environment}[1]{\color{red} }{}

\renewcommand{\vec}[1]{\ensuremath{\mathbf{#1}}}
\newcommand{\GJ}[1]{\ensuremath{#1_{\textrm{\tiny{}GJ}}}}    
\newcommand{\PC}[1]{\ensuremath{#1_\mathrm{pc}}}    
\newcommand{\LC}[1]{\ensuremath{#1_{\textrm{\tiny{}LC}}}}    
\newcommand{\NS}[1]{\ensuremath{#1_\textrm{\tiny{}NS}}}    
\newcommand{\NSarg}[2]{\ensuremath{#1_{\textrm{\tiny{}NS},#2}}}    
\newcommand{\Vvac}{\ensuremath{\Delta{V}_\textrm{vac}}}    
\newcommand{\DFT}[1]{\ensuremath{#1_\textrm{\tiny{}D}}}    
\newcommand{\DFTarg}[2]{\ensuremath{#1_{\textrm{\tiny{}D},#2}}}    

\newcommand{\B}{B0826$-$34}

\def\deg{{$^{\circ}$}}

\begin{document}

\submitted{The Astrophysical Journal, doi:10.1088/0004-637X/751/1/1}

\title{On plasma rotation and drifting subpulses in pulsars;\\
using aligned pulsar \B\ as a voltmeter.}

%
%
\author{J. van Leeuwen$^{* 1}$ and A. N. Timokhin$^{2,3,4}$}
\affil{
  $^1$Netherlands Institute for Radio Astronomy (ASTRON), Postbus 2,
  7990 AA Dwingeloo, The Netherlands, \small{leeuwen@astron.nl}\\
  $^2$NASA Postdoctoral Program Senior Fellow, Astrophysics Science
  Division, NASA/Goddard Space Flight Center,\\
  Greenbelt, MD 20771, USA, \small{andrey.timokhin@nasa.gov}\\
  $^3$Astronomy Department, University of California at Berkeley,
  Berkeley, CA 94720, USA\\
  $^4$Moscow State University, Sternberg Astronomical Institute,
  Universitetskij Pr. 13, 119992 Moscow, Russia\\
}

\date{Received 2012 January 16; accepted 2012 April 20; published 2012
June 7}

\begin{abstract}

We derive the exact drift velocity of plasma in the pulsar polar cap,
in contrast to the order-of-magnitude expressions presented by
Ruderman \& Sutherland (1975) and generally used throughout the
literature. We emphasize that the drift velocity depends {\em not on
  the absolute value}, as is generally used, but on the
\emph{variation} of the accelerating potential across the polar cap.
If we assume that drifting subpulses in pulsars are indeed due to this
plasma drift, several observed subpulse-drift phenomena that are
incompatible with the Ruderman \& Sutherland family of models can now
be explained: we show that variations of drift rate, outright drift
reversals, and the connection between drift rates and mode changes
have natural explanations within the frame of the ``standard'' pulsar
model, when derived exactly.  We apply this model for drifting
subpulses to the case of PSR\;\B, an aligned pulsar with two separate
subpulse-drift regions emitted at two different colatitudes. Careful
measurement of the changing and reversing drift rate in each band
independently sets limits on the
variation of the accelerating potential drop. The derived variation is
small, $\sim$$10^{-3}$ times the vacuum potential 
drop voltage.  We discuss the implications of this result for pulsar modeling.
\end{abstract}

\section{Introduction}
\label{Sect:Intro}
\makeatletter{\renewcommand*{\@makefnmark}{}\footnotetext[0]{$^*$Both authors contributed equally.}\makeatother}
Drifting subpulses are a modulation of the main pulsations seen in
radio pulsars.  Several months after the pulsar discovery announcement
by \citet{hbp+68}, this fast variation, or second periodicity, was
already reported. By this time, the original suggestion that the
pulsar's main periodicity originated from stars that pulsated competed
with models of neutron stars that rotated. Thus, the detection by
\citet{dc68} of the ``class two effect'', a much faster second
periodicity ($\sim$10\,ms) within the $\sim$1\,s main period, was
originally interpreted as a sign of a fast stellar pulsation,
modulated by the slower stellar rotation. Yet the amount of variation
seen within this second periodicity was larger than was generally
expected for neutron-star pulsations. In two papers in 1970, Don
Backer \nocite{bac70b, bac70c} showed that the ``marching subpulses'',
as they were by now called, change period throughout the main pulse
window, making stellar pulsations less likely but favoring
magnetospheric interpretations.

This pulsar magnetosphere is filled with dense plasma and the electric
field is shielded almost everywhere. Only in some geometrically small
regions it is not; there the electric field parallel to the magnetic
field is capable of accelerating charged particles.  Such is the
current radio-pulsar ``standard model'', often called the force-free
model, first introduced by \citet{GJ}, that we will refer to
throughout this paper. It agrees very well with observational
properties of pulsars: pulse peaks in both radio and gamma rays are
very narrow, indicating small emitting regions, and hence, small
regions of particle acceleration. In regions of closed magnetic field
lines the plasma is trapped; but plasma around open magnetic field
lines flows away, forming the pulsar wind, and needing constant
replenishment.

This dense plasma consists mainly of electrons and positrons
\citep{Sturrock71}, and is created in small regions with strong,
accelerating electric fields.  There is compelling observational
evidence of ongoing generation of electron-positron plasma: pulsars
stop emitting in radio roughly where their parameters drop below the
threshold for pair creation; and observations of Pulsar Wind Nebulae
indicate they are fed by flows of dense plasma.  The most plausible
place for pair creation is the region of open magnetic field lines in
the pulsar polar cap
\citep{RudermanSutherland1975,Arons/Scharlemann/78,Daugherty/Harding82}.
This might be not the only place of pair production in the pulsar
magnetosphere \citep[cf.\ ][]{Cheng/Ruderman76}, but without pair
production in the polar cap it is difficult to imagine how the
``standard model'' can work. Thus, the existence of an accelerating
region in the pulsar polar cap is an integral part of the
``standard model''.

Plasma fills the whole closed magnetic field-line zone and co-rotates
with the neutron star (NS).  Dense plasma in the open magnetic field
lines, however, exists only above the accelerating region: the region
with screened electric field where magnetic field lines are frozen
into the plasma is separated from the NS by an accelerating region
with a strong electric field. Now, the existence of this accelerating
region in general causes rotation of plasma relative to the NS.  As
pulsar emission is most likely generated by the plasma in the region
of open magnetic field, the power spectrum of that emission must have
a feature due to this plasma rotation relative to the NS.  We wish to
emphasize that this is a general statement which does not depend on
any particular model of pulsar radio emission.  If, more specifically,
``stable'' emitting features exist in the plasma, such as current
filaments or spark columns \citep{RudermanSutherland1975}, then these
will manifest as subpulses drifting across the pulse profile.

There are alternative models for the drifting subpulses, not
explicitly involving 
plasma rotation relative to the NS; a concise review can be
found in \citet{Kuijpers2009}.  Some of these
\citep[e.g.][]{Gogoberidze2005,Clemens2004} rely on some sort of
standing wave pattern; the beating of the wave period with the rotational
period of the pulsar provides a slow drift of emission features.  In such
models there are no distinct physical features actually moving
relative to the star.  However, fluctuation spectra for PSR\;B0943$+$10
have revealed 
symmetrical sidebands around its main phase modulation feature,
  indicative of an amplitude modulation every 20 drift bands
  \citep{DeshpandeRankin1999}; 
the most natural explanation is
the existence of actual long-lived emitting columns in the pulsar polar cap,
columns that rotate relative to the NS. The model of 
\citet{FungKhechinashviliKuijpers2006} suggests that a diochotron
instability due to differential rotation of plasma gives rise to plasma
columns in the open magnetic field line zone. These rotate relative to
the NS, but with an angular velocity that is generally different from the
angular velocity of the bulk plasma rotation.  Although 
the current-density profiles considered in
\citet{FungKhechinashviliKuijpers2006}  
are not directly related to existing pulsar models, in our opinion
the proposed mechanism should be studied in more detail in the future.

In this paper we consider the general idea that drifting subpulses are
caused by plasma drift relative to the NS.  Such plasma rotation in
the open field line region is an integral part of the standard model.  Our
goal is not to establish a detailed model for drifting subpulses, but
rather to explore testable predictions which can be made under such
general assumptions about the origin of subpulse drift.  We do not
specify a particular mechanism leading to the formation of emitting plasma
columns, but assume their existence causes regular drifting subpulses.  Even if
ultimately it turns out that drifting subpulses are not caused explicitly by plasma
rotation, this plasma rotation should still be visible as features in the power
spectrum of pulsar emission%
\footnote{Indeed most of the pulsars studied by
  \citet{wes06,Weltevrede2007} have features in the
  fluctuation (power) spectrum rather than clearly visible
  driftbands.}%
.

In Section~\ref{sec:drift_theory} we derive an expression for the
angular velocity of plasma rotation relative to the NS. We point out a
widespread misconception arising from the literal application of an
order-of-magnitude expression for this angular velocity from
\citet{RudermanSutherland1975}. In that Section we also argue that,
using the exact expression, effects such as drift
rate variations and drift direction reversals have a natural
explanation within the standard model.

In Section~\ref{sec:b} we introduce pulsar PSR\;\B\ (hereafter: \B), a
bright aligned rotator with subpulse-drift throughout the pulse
window.  In Section~\ref{Sect:Obs} we outline the observational set up
with which we measure such drift rate variations. In
Section~\ref{Sect:Modeling} we present the relations found in
Section~\ref{sec:drift_theory} in a form that can be directly applied
to such drift rate observations%
\footnote{Fitting code available at \\
  {\tt \scriptsize \url{http://www.astron.nl/pulsars/papers/lt12/}}}, 
and we describe the results from this fitting. We then derive the
maximum variations in potential drop over the polar cap of \B,
effectively using the subpulse drift in \B\ as a voltmeter. The
discussion and interpretations of these results take place in
Section~\ref{Sect:Discussion}.  We summarize and conclude in
Section~\ref{Sect:Conclusions}.

~\\

\section{Plasma rotation and drifting subpulses}
\label{sec:drift_theory}

\subsection{Plasma rotation in the open magnetic field line zone}
\label{sec:drift_theory_1}

\begin{figure}
  \begin{center}
    \includegraphics[width=\columnwidth]{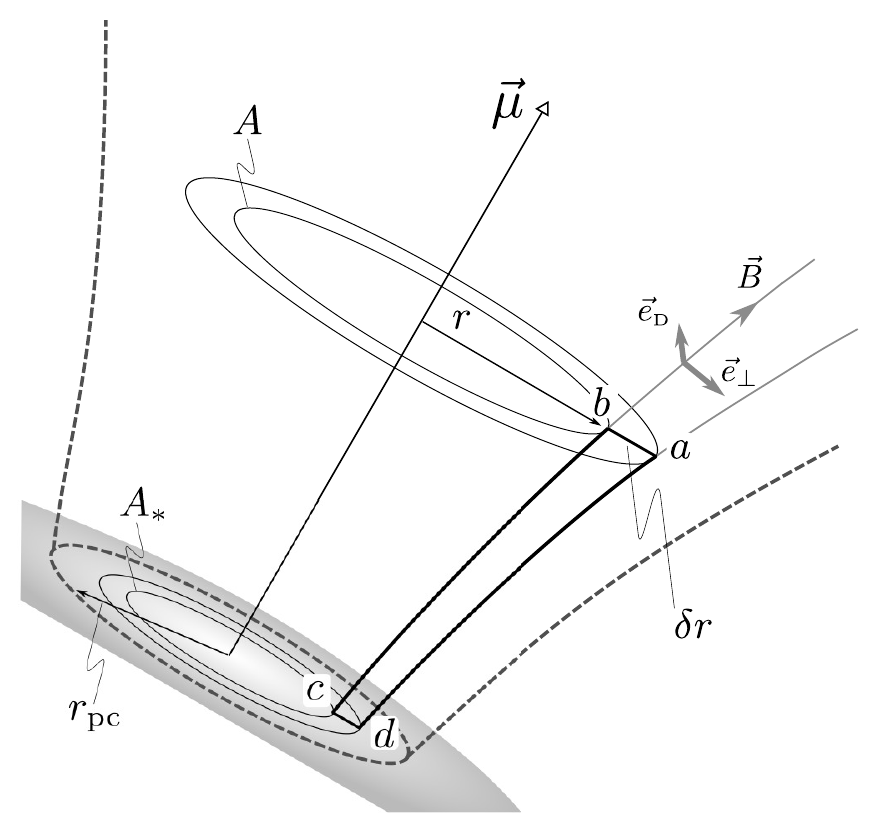}
  \end{center}
  \caption{Schematics of the pulsar polar cap. The polar cap
    boundary is shown with the dashed line. The path sections are
    explained in detail in the main text.}
  \label{fig:drift_rate_derivation}
\end{figure}

Let us do a more careful derivation of the formula for the drift speed
of plasma above the accelerating gap in the pulsar polar cap, than
that presented in \citet{RudermanSutherland1975}.  In the reference
frame corotating with the NS, Faraday's law for electric and
magnetic field as measured in the corotating frame has the same form as
in the laboratory frame \citep{Schiff1939}, namely
\begin{equation}
  \label{eq:Faraday_law}
  \nabla\times\vec{E} = \frac{1}{c}\frac{\partial{\vec{B}}}{\partial{t}}\,.
\end{equation}
The electric field in eq.~(\ref{eq:Faraday_law}) represents the
non-corotational component of the electric field -- the component
which forces plasma to move in the rotating frame.  In the corotating
frame the temporal variations of the magnetic field in the polar cap
can be caused only by fluctuations of currents in the magnetosphere.
Even if fluctuations of the electric current are of the order of the
Goldreich-Julian current $\GJ{\rho}c$, the resulting variation of the
magnetic field are
$\delta{B}/B\simeq{}(\NS{R}/\LC{R})^{3/2}\simeq{}6\times10^{-6}P^{-3/2}$;
here $\NS{R}$ is the NS radius and $\LC{R}\equiv{}c/\Omega$ is the
Light Cylinder radius, $\NS{\Omega}$ is the angular velocity of NS
rotation, $P$ is pulsar period in seconds.  Hence
$\nabla\times\vec{E}=0$ with high accuracy and circulation of the
non-corotational electric field along a closed path is zero (even for
an inclined rotator).

Let us consider the circulation of the electric field along a fiducial
path $abcd$ -- see Fig.~\ref{fig:drift_rate_derivation} -- in the
polar cap of pulsar the top ($ab$) and bottom ($cd$) parts are
infinitesimally small and are just below the NS surface ($cd$) and
above the accelerating gap in the region of dense plasma with screened
electric field ($ab$); the lateral sides ($bc$ and $da$) follow
magnetic field lines.  Note that we consider a path which is fully in
the open magnetic field line region, in contrast to the work of
\citet{RudermanSutherland1975} where one lateral path was along the
field line along the polar cap boundary (see their Fig.~4).  To
  clarify our derivation, Fig.~\ref{fig:drift_rate_derivation} shows
  our unit vectors along positive directions for the perpendicular
  component of the electric field, $e_\perp$, and the drift velocity,
  $e_D$. The circulation of the electric field along this path is
\begin{eqnarray}
  \label{eq:int_E_dl}
  \oint{}E\,dl & = & 
  \displaystyle
  - E_\perp \delta{}r + \int_b^cE_{||}\,dz + \int_d^aE_{||}\,dz
  \nonumber\\
  & = & 
  \displaystyle
  - E_\perp dr + V_{bc} - V_{ad}= 0\,,
\end{eqnarray}
where $E_\perp$ and $E_{||}$ are components of the electric field,
perpendicular and parallel to the magnetic field respectively, 
while
$V_{bc}, V_{ad}$ are potential drops in the accelerating gap, along
magnetic field lines.  $E_\perp=0$ just below the NS
surface which we assume to be a perfect conductor.  Taking the limit
$\delta{}r\rightarrow0$ we get for the non-corotational component of
the electric field
\begin{equation}
  \label{eq:E_perp}
  E_\perp = -\frac{dV}{dr}\,.
\end{equation}
The drift velocity of the plasma relative to the NS is then
\begin{equation}
  \label{eq:v_drift}
  \DFT{v} = c\frac{\vec{E}\times{}\vec{B}}{B^2} = \frac{c}{B} \frac{dV}{dr}\,.
\end{equation}
The drift velocity of plasma relative to the NS thus depends on the
\emph{variation} of the electric potential in the acceleration zone.
 For aligned pulsars this result is well-known textbook material
  \citep[e.g.][]{BeskinBook2009} and for inclined rotators the correct
  general expression for the plasma drift velocity was explicitly
  mentioned in \citet{FungKhechinashviliKuijpers2006}; as we shall
  detail below, this result partially supersedes the Ruderman \&
  Sutherland family of models, but it hasn't yet struck root. Here we
  want to draw special attention to it, as it has profound
  consequences for the interpretation of pulsar drift phenomena.

The reasoning leading up to this conclusion, i.e.\ the \emph{variation}
of the accelerating potential causes plasma drift, can be put
differently and perhaps more insightful: the electric field is the
gradient of the electrostatic potential, $\vec{E}=-\nabla{}V$; in the
co-rotating frame the potential is the accelerating potential and the
perpendicular electric field $E_\perp$ is the field which causes drift
relative to the NS; hence,
$E_\perp=-(\nabla{}V)_\perp=-\partial{}V/\partial{}r$. If, for example,
the potential drop is the same along all polar-cap magnetic field
lines, there is no relative rotation of plasma, however large the
potential drop is. In that case though, there is a region very close
to the polar cap boundary where the potential drop plummets to zero at
the last closed magnetic field line. Plasma in that region will have a
very large drift velocity.

The original derivation of the expression for $\DFT{v}$ in
\citet{RudermanSutherland1975} is similar to our derivation here, but
instead of expressing $\DFT{v}$ in terms of the derivative of the
accelerating potential $dV/dr$, an order-of-magnitude estimate is
used: their eq~(31) is equivalent to our eq.~(\ref{eq:v_drift}), but
using the entire potential drop along a field line $V$ for $dV$, and
half of the polar cap radius, $\PC{r}/2$, as an estimate for $dr$.
Yet, the potential drop in the pair-production zone is expected not to
  vary much from one field line to another, except at the boundaries
  of the active zone
  \citep[e.g.][]{Harding/Muslimov98,Hibschman/Arons:pair_multipl::2001};
  the active zone is also expected to be a noticeable part of the polar
  cap, at least in the space charge limited flow model
  \citep{Arons1979}.  Hence, the \citet{RudermanSutherland1975}
  order-of-magnitude prediction for the drift rate is a large
  overestimate.  And indeed, the drift rate speeds
  predicted by \citet{RudermanSutherland1975} eq.~(31) are much larger
  than those observed -- for example, the very slow
  but steady drift rate observed in PSR\;B0809+74 \citep{lsr+03} is
  incompatible with this order-of-magnitude estimate.  Thus, in
the most detailed drifting-subpulse models based on
\citet{RudermanSutherland1975}, a substantial reduction of the
accelerating potential is necessary to reproduce these observed slow
drift rates. In \citet{GilSendyk2000} and
\citet{GilMelikidzeGeppert2003}, this is achieved by assuming highly
non-dipolar polar-cap magnetic fields, and by fine-tuning the particle
flux impact to produce the critical surface temperature.

Furthermore, literal application of Ruderman \& Sutherland's
(\citeyear{RudermanSutherland1975}) expression for the plasma drift
speed (i.e.\ in the form $\DFT{v}=cV/\PC{r}B$) has difficulty
explaining the drift-rate variations with time or longitude that are
observed in several pulsars \citep[such as the bidirectional drift
  bands in PSR\;J0815+09,][]{clm+05}, and is incapable of producing
the complete drift direction reversals seen in PSR\;\B.
The explanation for such observed subpulse-drift reversals usually involves
aliasing. In the case of \B\ this requires 
the circulation frequency to be fine-tuned to the rotation frequency,
such that the aliased subpulses appear to move only very slowly; and
indeed \citet{ggks04} were able to construct such a model.
However, \citet{elg+05} find that the observed reversals
influence the intrinsic subpulse brightness, strongly suggesting that
the reversal is not just apparent, but actually occurs in the
corotating frame. The existence of such
real reversals thus suggests a failing of the $\vec{E}\times\vec{B}$ family of
  models.

We want to point out that all these apparent discrepancies between
model and observation are non-existent. Using the derivation we
present here, in eq.~(\ref{eq:v_drift}), no fine-tuning or additional
assumptions, such as strong higher-order magnetic fields, are needed
to explain the observed subpulse-drift properties. The drift rate
depends on the variation of the potential drop from one field
line to another.  The potential drop can be very large, but if it does
not change much across the polar cap the drift rate will be slow.
Depending on whether the potential drop increases or decreases toward
the center of the polar cap the drift rate will be in one or another
direction. If the potential-drop derivative at the line of sight
varies with time, the observed drift speed changes, and can even
reverse. In Section~\ref{Sect:Modeling}, we use subpulse-drift data
from \B\ to show that the variation of the accelerating potential
necessary to explain these phenomena is small indeed, and that the
model is plausible.

We reiterate that we are not solving these problems by introducing
extra assumptions on how the potential drop
varies. Eq.~(\ref{eq:v_drift}) simply \emph{is} the correct dependence
for the drift velocity of the plasma, it is not a fine-tuning or
modification of the previously widely used order-of-magnitude
expression for $\DFT{v}$. If drifting subpulses are indeed due to
plasma motion relative to the NS, all their properties depend on the
variation of the potential drop and not on the potential drop itself.
In Section~\ref{Sect:Discussion} we will also argue that all the
properties of the accelerating potential necessary for explanation of
various properties of drifting subpulses seem to be integral
properties of the standard pulsar model.

\subsection{Connecting with observations of drifting subpulses}
\label{sec:observational_quantities}

In pulsars with subpulse drift, eq.~(\ref{eq:v_drift}) can be used to
set limits on the local parameters of the pulsar polar cap. There, a
measurement of the plasma drift velocity provides the value of the
radial derivative of accelerating potential.  In practice however, the
quantity that subpulse-drift observations produce is the angular drift
velocity of subpulses within the pulsar profile, not the plasma drift
velocity in the pulsar coordinate frame that is needed in
eq.~(\ref{eq:v_drift}). Below we derive  expressions that connect
observable parameters to the variation of the accelerating potential.

The momentary
angular velocity of plasma relative to a magnetic axis is
\begin{equation}
  \label{eq:delta_omega_dVdr}
  \DFT{\Omega} = \frac{c}{B r} \frac{dV}{dr}\,,
\end{equation}
where $r$ is the distance between our sight-line path and the magnetic
axis.  When we derived
this expression in Section~\ref{sec:drift_theory_1}, we considered the
path segment $ab$ in the region 
above the accelerating gap where electric field is screened, and so
$B,r,dr$ in eq.~(\ref{eq:delta_omega_dVdr}) are also taken in some plane
perpendicular to the magnetic axis $\boldsymbol\mu$, in the region with
screened electric field (see Figure \ref{fig:drift_rate_derivation}).

Our path $abcd$ has lateral sides that follow magnetic field lines,
hence if $\Phi$ is the magnetic flux limited by some closed path
passing through our fiducial point (e.g. the path $A$ or $A_*$ in
Fig.~\ref{fig:drift_rate_derivation}) then
$Br=\partial{\Phi}/\partial{r}\partial{\varphi}$, where $\varphi$ is
the azimuthal angle.  For any magnetic field that is axisymmetric
relative to some magnetic axis, the plasma drift relative to
the NS will be a rotation around this magnetic axis.  The magnetic
flux can be measured as the flux trough a circle of radius $r$, and
equation (\ref{eq:delta_omega_dVdr}) takes the form
\begin{equation}
  \label{eq:OmegaD__dVdPhi}
  \DFT{\Omega} = 2 \pi{}c\,\frac{dV}{d\Phi}\,.
\end{equation}

It will be convenient to normalize the potential drop along a
  magnetic field line to the potential drop across magnetic field
  lines at the NS surface between the rotation axis and the boundary
  of the polar cap in an aligned pulsar with the same period:
\begin{eqnarray}
  \label{eq:Vvac}
  \Vvac & = & 
  \displaystyle
  - \int_0^{\PC{r}} E\,dr = 
  \frac{1}{2}\frac{\NS{\Omega}}{c} B \PC{r}^2 = 
  \frac{\NS{\Omega}\PC{\Phi}}{2\pi{}c}\nonumber\\
  & \simeq & 
  \displaystyle
  6.6\times10^{12}\NSarg{R}{6}^3 B_{12} P^{-2} \; \mbox{Volts}\,,
\end{eqnarray}
here $\PC{r}$ is the polar cap radius, $\PC{\Phi}$ is the magnetic
flux through the polar cap, $B_{12}$ magnetic field strength in units
of $10^{12}$~G, and $\NSarg{R}{6}$ is the NS radius in $10^6$~cm.  In
doing this numerical estimate we assumed that the polar cap is small
enough that the variation of the magnetic field over it is negligible.
$\Vvac$ is a good estimate for the available potential drop -- if
  there were vacuum in the open magnetic field line zone the potential
  drop along magnetic field lines in the polar cap would reach a value
  comparable to $\Vvac$ at the height $\sim\PC{r}$ from the NS surface
  and then vary very slow with the altitude, approaching its maximum
  value $\Vvac$ from below.

Normalizing the magnetic flux to $\PC{\Phi}$ we get from
eq.~(\ref{eq:OmegaD__dVdPhi})
\begin{equation}
  \label{eq:OmegaD_OmegaNS}
  \frac{\DFT{\Omega}}{\NS{\Omega}} = \frac{d\tilde{V}}{d\tilde{\Phi}}
\end{equation}
where $\tilde{V}\equiv{}V/\Vvac$ and
$\tilde{\Phi}\equiv{}\Phi/\PC{\Phi}$. The normalized magnetic flux is
$\tilde{\Phi}=(r/\PC{r})^2= (\theta/\PC{\theta})^2$, where $\theta$ is
the colatitude measured from the magnetic axis.  Let us denote the
colatitude to the colatitude of the polar cap boundary as
\begin{equation}
  \label{eq:xi}
  \xi \equiv \frac{\theta}{\PC{\theta}}\,,
\end{equation}
then in terms of $\xi$ eq.~(\ref{eq:OmegaD_OmegaNS}) takes the
form
\begin{equation}
  \label{eq:OmegaD_OmegaNS__xi}
  \frac{\DFT{\Omega}}{\NS{\Omega}} = \frac{1}{2\xi}\frac{d\tilde{V}}{d\xi}  
\end{equation}

After one rotation the difference in phase between a fiducial point at
the NS surface and an emitting column will be
\begin{equation}
  \label{eq:d_phase}
  \delta{\phi}_1 = \DFT{\Omega} P =
  \frac{\DFT{\omega}\pi}{180^\circ}\,,
\end{equation}
where the $\DFT{\omega}$ is the drift velocity in units of
degrees/period.  Using eq.~(\ref{eq:OmegaD_OmegaNS__xi}) we can then write
the derivative of the accelerating potential as
\begin{equation}
  \label{eq:vd__dVdxi}
  \frac{dV}{d\xi} = \xi\frac{\DFT{\omega}}{180^\circ}
\end{equation}

Thus, if we measure the angular velocity at one colatitude, we can now
calculate the derivative of the accelerating potential at that point.
If we could measure it at two different colatitudes, we can set
limits on the accelerating potential, assuming a gradual variation of
the potential across the polar cap.  For such a study our line of
sight must cross two separate drift bands at different colatitudes. An
ideal case would be a bright, slightly inclined rotator where the line
of sight crosses two regions at different colatitudes, each for a
significant fraction of period such that any drifting subpulses are
clearly visible; \B\ is such a pulsar.  It has two separate drift
bands which are thought to represent emission from different
colatitudes.  This pulsar provides us an unique opportunity to put
direct observational constrains on the physical conditions in the
polar cap accelerating zone -- the heart of the pulsar.

\section{PSR\;\B\ }
\label{sec:b}

\begin{figure}
  \includegraphics[width=0.45\textwidth]{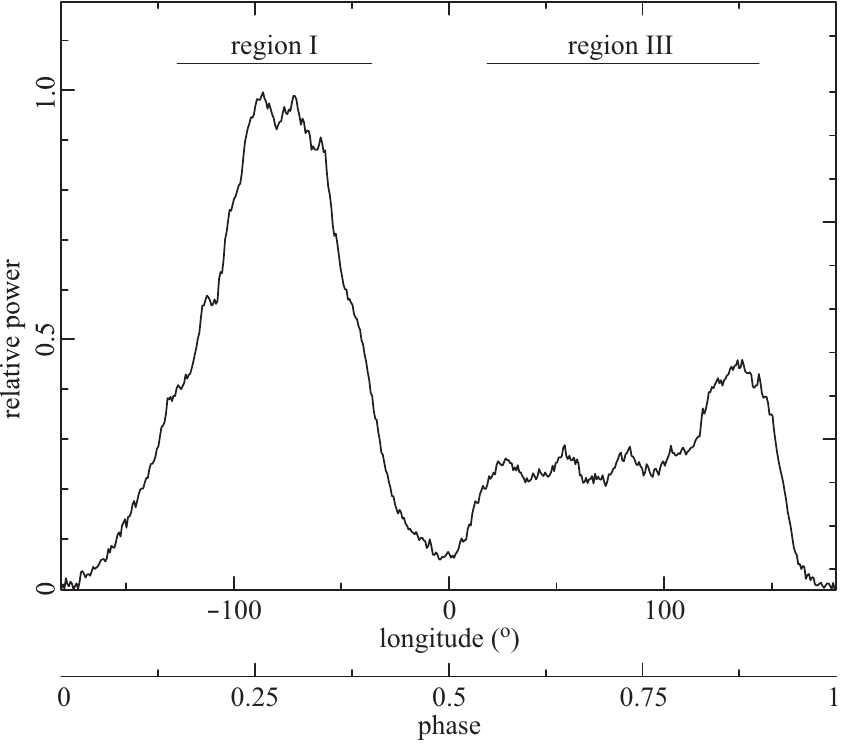}
\caption[]{ The  360\,$^\circ$ average profile of \B\ in the strong
  mode. 1500 seconds of data from observation PT0168\_116 were integrated. }
\label{Fig:Profile}
\end{figure}

Pulsar \B\ was discovered as a broad-profile but very intermittent
pulsar \citep{mlt+78}. Figure \ref{Fig:Profile} shows how the pulse
profile basically spans 360\deg. \B\  emits strongly only 30\% of the
time. When not in this strong mode, no emission could be detected and
the pulsar appeared to be in a ``null'' state \citep{dll+79}. 
Several years after these
first studies at 408 and 635\,MHz respectively, \citet{elg+05} revisit
\B\ with new Parkes data at 1.4\,GHz, and find that during the null
state a weak, different pulse profile is emitted (cf. Figure
\ref{Fig:OnOff}). In \citeyear{bgg08}, \citeauthor{bgg08} do not
detect a weak-mode profile in GMRT data at six frequencies between 157
and 1060\,MHz; yet Parkes observations at 685 and 3094\,MHz confirm
the existence of this weak mode emission \citep{sery11}.

Soon after the initial discovery, \B\ was found to display drifting
subpulses over more than half of its pulse period. At times, up to 9
drift tracks are visible. Most remarkable is the occasional 
reversal of drift direction \citep{bmh+85}.

The combination seen in {\B}, of a broad profile that cuts through
different lines of sight from the magnetic pole, and a system of
highly varying drifting subpulses, is unique.  The number of
pulsars that show some kind of periodic modulation is substantial
\citep[$>$50\%, see][ for an overview]{wes06}; and most of those show
drift-rate variations with time and/or pulse longitude.  But there is
only a handful, at most, of bright wide-profile or nearly-aligned rotators
like \B\ known \citep[e.g.\ B0818$-$41,][]{bggs07}.
And although some pulsars, such as PSR\;B2303+30 \citep{rwr05}, have
different modes in which drift rates appear to be in opposite
directions, \B\ is the only source in which the intrinsic drift rate
gradually reverses outright \citep{elg+05}.

\begin{figure}
  \includegraphics[width=0.45\textwidth]{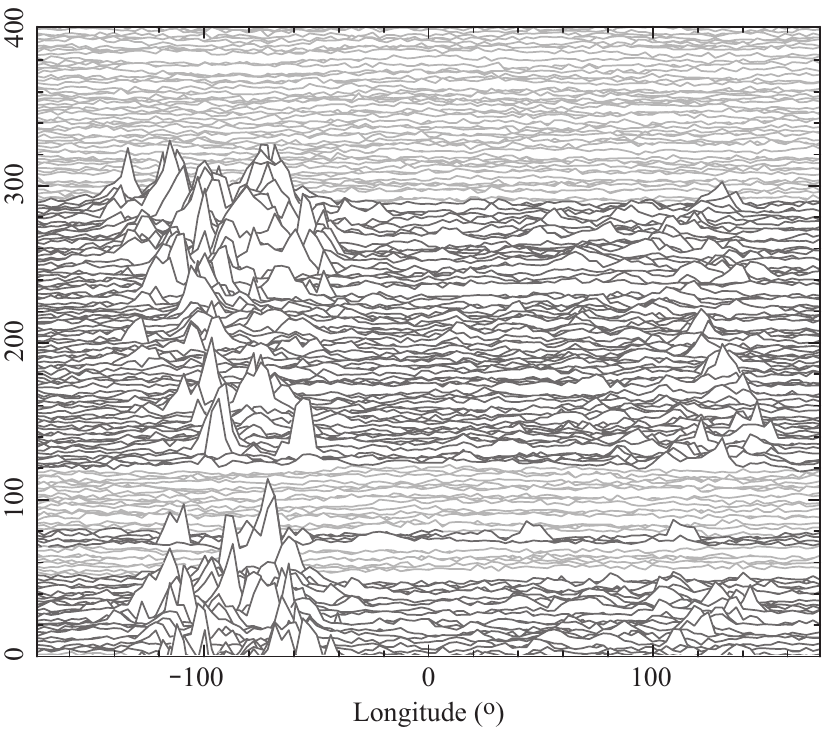}
\caption[]{ A pulse stack with several nulls in light gray and a
  transition to a weak mode interval lasting several thousand pulses, starting at
  pulse number 300. For both the nulls and the mode transition, there
  is a clear separation between on and off pulses in the pulse-energy
  histogram. }
\label{Fig:transitionStacked}
\end{figure}

\section{Observations and data reduction}\label{Sect:Obs}
Pulsar \B\ was observed for a total of 20\,hrs on 2002 Sep 10 and 11, and
Nov 1 and 2. The middle beam of the multibeam receiver on the Parkes
telescope was used, at a central observing frequency of 1372 MHz.  A
288\,MHz wide observing band was split over 96 $\times$ 3 MHz channels
in the analog filter bank \citep{mlc+01}. The total power from each of the
filter channels every was recorded every 0.25/1.0\,ms (Table
\ref{Table:Obs}).

\begin{table}
\begin{center}
\begin{tabular}{l|rrrrrl}
OBSID       & t      &  Date               & $t_{samp}$  &  Strong mode \\
            & (hr)   &  (MJD)              & (ms)      &    (s)         \\
\hline
PT0168\_116 &      6 & 52527             &       1.0   &    1939$-$3535 \\ 
PT0168\_150 &      6 & 52528             &       1.0   &    (none)     \\ 
PT0169\_106 &      4 & 52579             &       0.25  &    0$-$7221 \\
PT0169\_152 &      4 & 52580             &       0.25  &    (none)  \\

\end{tabular}
\caption{
Details of the four observing sessions used in this paper. Observation
identifier $OBSID$, duration $t$, date, and sampling time $t_{samp}$ are
shown. The last column marks the begin and end times of any strong
mode sections in the data, indicated in seconds since the start of the
observation. }
\label{Table:Obs}
\end{center}
\end{table}

These filter bank data were retrieved from the {\tt PT\_TAPES} section
of the Australian Pulsar Timing Archive\footnote{\tt
  http://www.atnf.csiro.au/research/pulsar/archive/}, for projects
{\tt P276} and {\tt P417}.  They were unpacked using {\tt sc\_td} and
converted from the original 1\,bit to the more standard 8\,bit filterbank format
for compatibility\footnote{\tt
  http://sigproc.sourceforge.net/}. We next inspected each of the four
observations with PRESTO\footnote{\tt
  http://www.cv.nrao.edu/$\sim$sransom/presto/} to remove radio
  interference.  Starting 
from the known ephemeris \citep{hlk+04} and the dispersion measure of
52 pc/cm$^3$ \citep{bgg08}, the data were folded and dedispersed at
the period, period derivative and dispersion measure that maximized
signal to noise ratio.

\subsection{Identifying mode changes and nulls}
\begin{figure}[b]
  \begin{center}
  \includegraphics[width=0.45\textwidth]{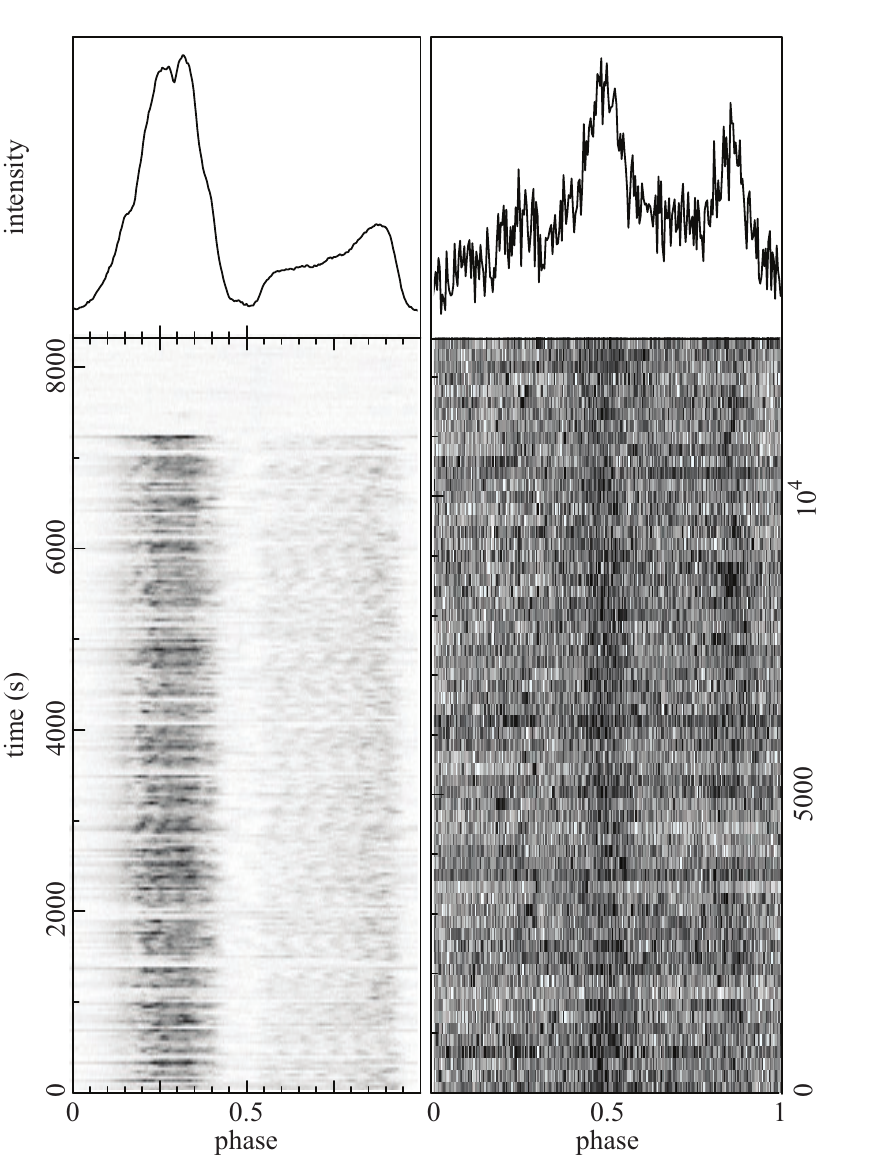}
  \end{center}
\caption[]{Plots of intensity in gray scale versus phase and time
  (bottom panels), and of the integrated profiles over the entire
  observation (top panels). The intensity in the top panels is in
  arbitrary units, and the two observations are not plotted to scale.
  A 2-hr strong-mode sequence from the start of observation
  PT0169\_106 is shown on the left, and a 4-hr weak-mode observation is shown on the right
  (PT0169\_152). }
\label{Fig:OnOff}
\end{figure}
\begin{figure*}[t]
  \includegraphics[width=\textwidth]{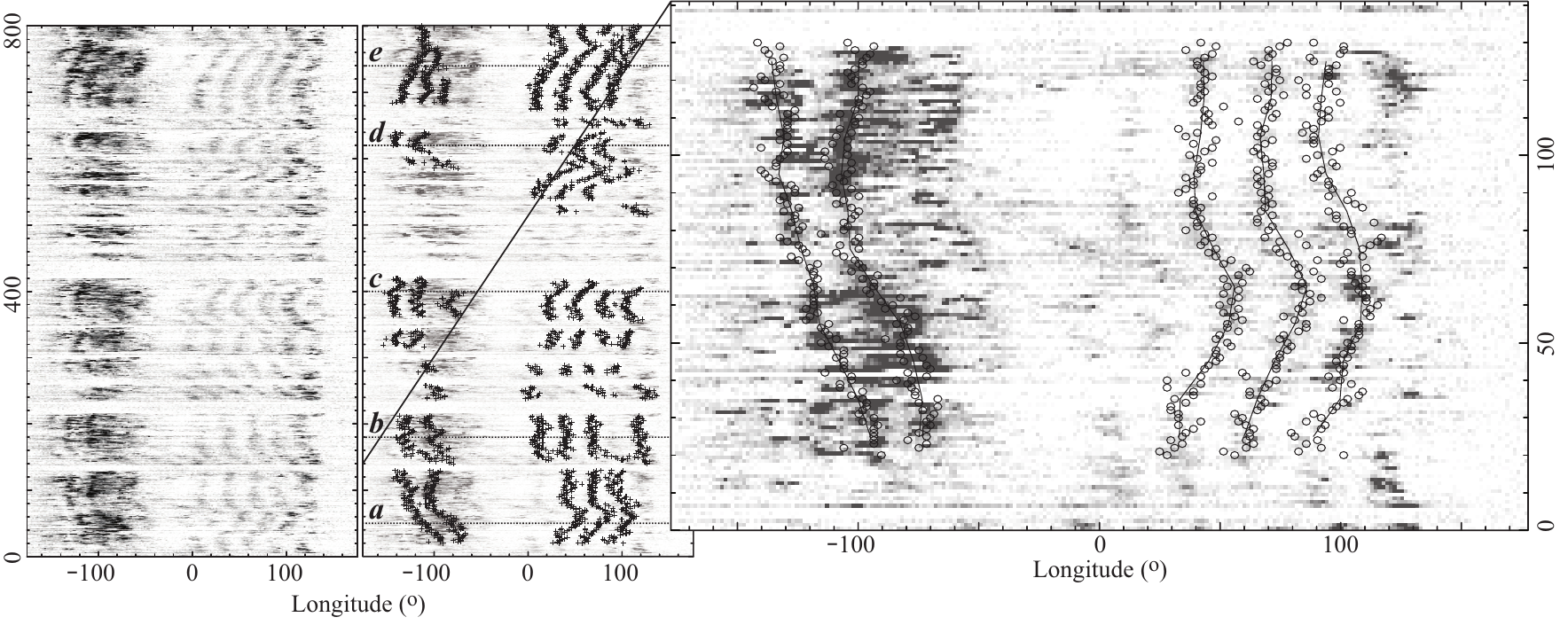}
\caption[]{ A strong-mode sequence about 800 pulses long from
  observation PT0168\_116 (left), the results of fitting for the
  locations of the subpulses (middle), and the final fits to drift
  tracks for the initial $\sim$100 pulses (right). Up to 5 or 6 drift
  bands are usually identified.  In the middle and right plot, the
  contrast of the pulse stack was halved to make the fit markers more
  visible.  For the dotted lines labeled $a$-$e$ in the middle panel
  the potential drop is derived and plotted in Fig.~\ref{Fig:V}. }
\label{Fig:FitBand}
\end{figure*}
Using fitSubPulses \citep{lkr+02} the data were next automatically and
visually inspected for nulls and mode changes. These are 
robustly separated in a pulse-energy histogram \citep{jl04}
and verified manually. In Figure
\ref{Fig:transitionStacked} we show several nulls, and a transition
from strong to weak mode. In two of the four sessions, {\tt
  PT0168\_150} and {\tt PT0169\_152} (Table \ref{Table:Obs}), the
pulsar emitted in the weak mode for the entire session. In the two
other observations strong-mode sequences are found, one of $\sim$2000
and one of $\sim$7000 seconds duration (i.e.\ $\sim$1000 and
$\sim$4000 individual pulses for this 1.85\,s period pulsar).  In
Figure \ref{Fig:OnOff} we show the long-term behavior in the strong
and weak mode, and a transition between the two.

\subsection{Determining drift rates}
\label{sec:detdrift}

Within the strong-mode sequences, we next worked toward identifying
drift rates and directions. Our aim was to characterize the variations
in subpulse drifting over the wide profile. We thus fit for the
location of individual subpulses. This is in contrast to the method
described by \citet{elg+05} for \B, where a comb-like template with 9
main drift tracks at fixed spacings was used to come to an estimate of
the driftrate averaged over all drift tracks. Such a method cannot
measure drift rates that vary with pulse longitude, as is seen in many
pulsars \citep{wes06}. For each strong-mode pulse, we thus fitted for
the location of individual subpulses: we smoothed the single-pulse
profile by a 4\,ms window (the approximate average width of the
subpulses) and then determined the location of the maximum. The
maximums in subsequent single pulses were next grouped together, if
they were less than half the average subpulse-separation apart in phase. Only tracks
with more than 10 subpulses were retained. In this manner, subpulse
tracks of more than a hundred pulses were robustly identified.  Within
each of these tracks, the variation in drift speed and direction was
next measured by least-squares fitting to subsequent sets of 10
subpulses, as illustrated in the right-most panel of Figure
\ref{Fig:FitBand}.

\section{Modeling}
\label{Sect:Modeling}

In the strong mode, \B\ emits mainly in two regions.  In Figures
\ref{Fig:Profile} and \ref{Fig:OnOff} these are the brighter phase
range 0.15$-$0.40 and the dimmer range 0.55$-$0.9. \citet{elg+05}
label these region I and III, respectively. They find that in region I
both the width and separation of the subpulses are reduced by a factor
1.22 compared to region III, from which they conclude that region I is
closer to the magnetic axis than region III.

\begin{figure}
  \includegraphics[width=0.45\textwidth]{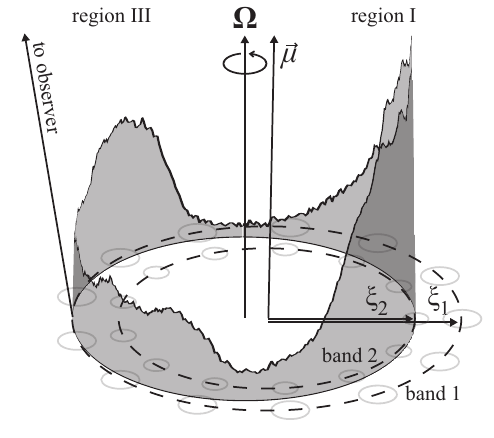}
\caption[]{ Diagram of the geometry used in this paper, for the polar
  cap region of \B. 
  Folded on the line-of-sight to the observer is the 360\deg pulse profile
  from Figure \ref{Fig:Profile}. Regions I and III from that Figure are
  also indicated. The angles between
  rotation axis  $\boldsymbol\Omega$, the magnetic axis
  $\boldsymbol\mu$, and the line of sight angle to the observer are
  from \citet{elg+05}; the system of two rings of
  subpulses, one realization of which is illustrated on the bottom
  plane, is after Figure 10 in that same paper. The colatitudes of our
  bands 1 and 2 are marked $\xi_{1,2}$.
}
\label{Fig:Geometry}
\end{figure}

We now use the drift speeds in these two regions to model the
variation of the acceleration potential in \B. We define band 1 as
the band near the edge of the polar cap, i.e.\ the region empirically
numbered III; and band 2 as being the closest to the magnetic axis, i.e.\ region I (the
left-most band in Figure \ref{Fig:OnOff}). The drift speeds and
colatitudes of band 1 and 2 are $\DFTarg{\omega}{1,2}$ and
$\xi_{1,2}$. We know the separation between these drift bands
$\Delta\xi\simeq0.2$ and assume band 1 is near the edge of the polar
cap, so $\xi_{1}\simeq1.0$ \citep{elg+05}.  In Figure 
\ref{Fig:Geometry} the observed strong-mode profile from Figure
\ref{Fig:Profile} is shown in this geometry. We have freedom for
setting $V$ at $\xi_1$ to an arbitrary value, let it be zero -- we
have an equation only for $dV/d\xi$. We thus have 3 known parameters --
$(dV/d\xi)_{1,2}$ and $V(\xi_1)=0$, such that we can fit for $V(\xi)$ with a
parabola:

\begin{equation}
  \label{eq:V_parabola}
  V(\xi) = a(\xi^2-\xi_1^2) + b(\xi-\xi_1)\,.
\end{equation}
Using this parabolic fit, we next derive the maximum potential drop
that our line of sight comes across in \B.
The values of the coefficients are determined from
eq.~(\ref{eq:vd__dVdxi}) at points $\xi_1,\xi_2$.  Depending on the
values of these coefficients, $V(\xi)$ either has a local extremum at the
point $\xi_{ex}\in[\xi_2,\xi_1]$, or it is a monotonic function of
$\xi$.  The position of the extremum is $\xi_{ex}=-b/2a$ and if it is
inside the interval $[\xi_2,\xi_1]$ the maximum potential drop
variation $\Delta{V}_{\max}^{ex}$ between points $1$ and $2$ is either
$V(\xi_{ex})$ or $V(\xi_{ex})-V(\xi_2)$, depending on what absolute
value is larger.  The general expression for the maximum potential
drop variation in colatitude interval $[\xi_2,\xi_1]$ is then
\begin{equation}
  \label{eq:Vmax}
  \Delta{V}_{\max} = 
  \left\{
    \begin{array}{ll}
      \Delta{V}_{\max}^{ex}, &
      \mathrm{if~} \xi_{ex}\in[\xi_2,\xi_1] \\
      V(\xi_2), &
      \mathrm{if~} \xi_{ex}\not\in[\xi_2,\xi_1]
    \end{array}
  \right.
\end{equation}

\begin{figure}
  \includegraphics[width=0.45\textwidth]{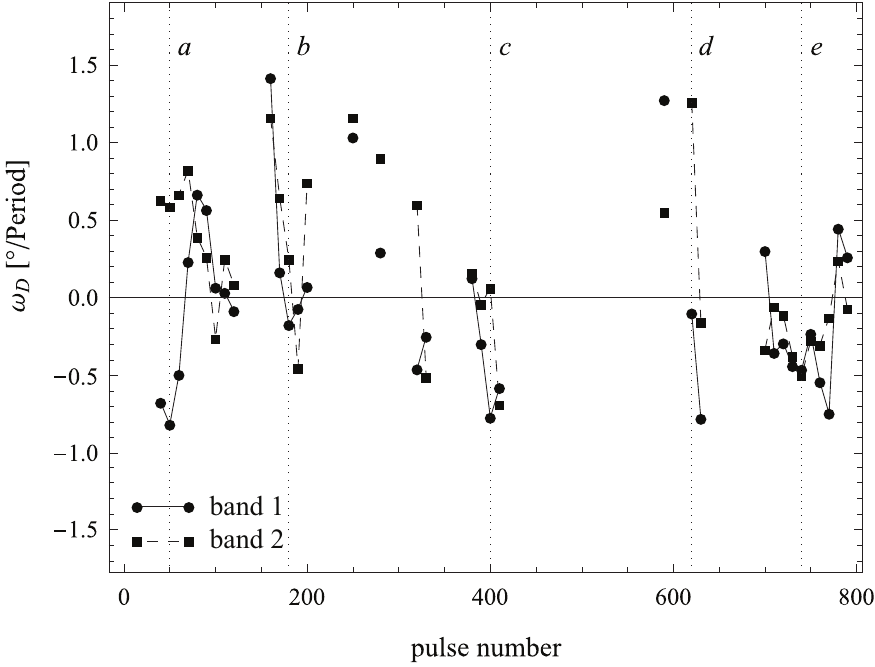}
  \caption[]{Drift speed $\DFT{\omega}$ (degrees per period) in two
    bands as functions of pulse number.  The drift speed for band 1 is
    shown by filled circles connected by a solid line, for band 2
    by filled squares connected by a dashed line. Dotted lines
    labeled $a-e$ show pulses for which we plot the potential drop in
    Fig.~\ref{Fig:V}. }
\label{Fig:DriftSpeeds}
\end{figure}

To each of the drift tracks identified in data set PT0168\_116, line
segments of 10 pulses were fitted, as 
shown in Figure \ref{Fig:FitBand}. The slope of these segments is
the drift velocity $\DFT{\omega}$, where we define
positive $\DFT{\omega}$ as drifting toward earlier arrival. 
For band 1, the drift rate of up to three tracks was averaged to obtain
$\DFTarg{\omega}{1}$, for band 2, $\DFTarg{\omega}{2}$ is the average
of up to two drift tracks (see Figure \ref{Fig:FitBand}). The
resulting drift speeds range from $-1.0$ to $+1.5$ degrees per period,
and are plotted in Figure \ref{Fig:DriftSpeeds}. Figures
\ref{Fig:FitBand} and  \ref{Fig:DriftSpeeds} show
that while bands 1 and 2 generally show the same drift rate, as
assumed for  data reduction in \citet{elg+05}, there are several
periods in which the drift rate between the bands is significantly
different.

\begin{figure}
  \includegraphics[width=0.45\textwidth]{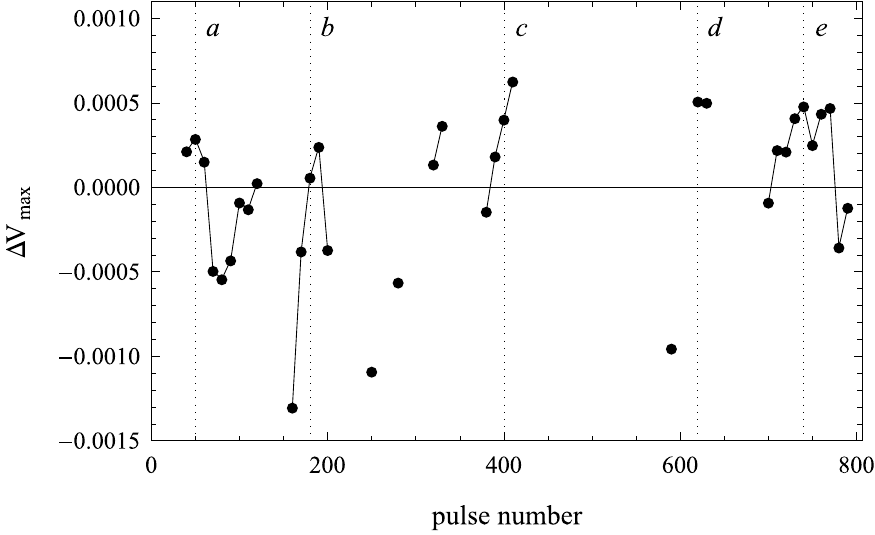}
  \caption[]{Maximum potential drop difference per eq.~(\ref{eq:Vmax})
    in the colatitude interval between the two drift bands, as a function of
    pulse number. $\Delta{}V_{\max}$ is normalized to the vacuum
    potential drop. Notations are the same as in
    Fig.~\ref{Fig:DriftSpeeds}.}
\label{Fig:Vmax}
\end{figure}

\begin{figure}
  \includegraphics[width=0.45\textwidth]{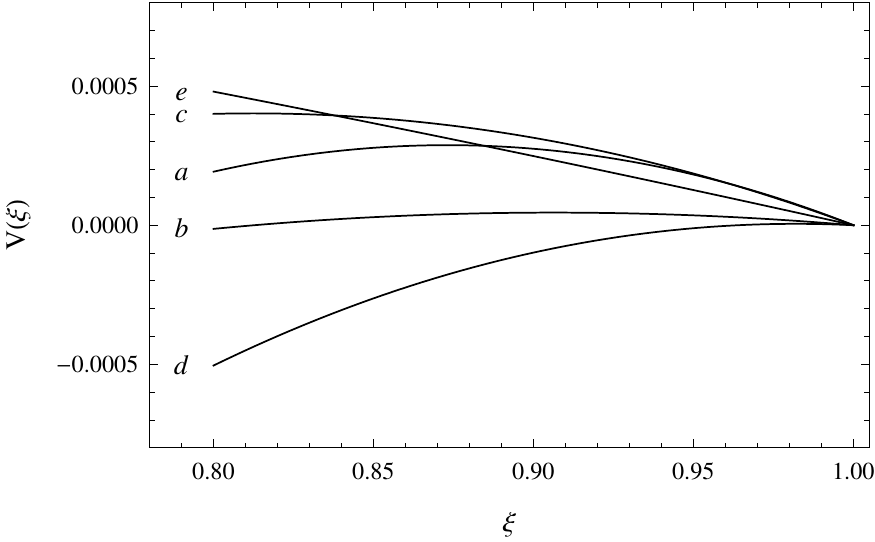}
  \caption[]{Potential drop variation according to
    eq.~(\ref{eq:V_parabola}) in the colatitude interval between the two drift
    bands, for pulses marked by $a-e$ on
    Figs.~\ref{Fig:DriftSpeeds} and \ref{Fig:Vmax}, as a function of the
    normalized colatitude $\xi\equiv\theta/\PC{\theta}$.  $V(\xi)$ is
    normalized to the vacuum potential drop.}
\label{Fig:V}
\end{figure}

For all sections with drift rate estimates, the maximum potential drop
difference $\Delta{}V_{\max}$ was next derived using
Eqs.~(\ref{eq:vd__dVdxi}--\ref{eq:Vmax}). The resulting potential drop
variations, plotted in Figure \ref{Fig:Vmax}, are of order of several
$10^{-4}\,{V}_\textrm{vac}$.  In Figure \ref{Fig:V} the fitted shape
of the potential drop is shown for five representative sections
labeled (a-e) in Figure \ref{Fig:FitBand}.

The potential drop variations shown in Figures \ref{Fig:Vmax} and
\ref{Fig:V} assume that our outer sight line traverse is close to the
edge of the polar cap, i.e.\ $\xi_{1}\simeq1.0$. If the actual geometry
is described with a smaller value for $\xi_{1}$, the reported potential
drop values scale down linearly.

A Mathematica notebook containing these equations as well as routines
to derive potential drop curves from driftrate data, is publicly
available\footnote{\tt \url{http://www.astron.nl/pulsars/papers/lt12/}}.

So, from Figures \ref{Fig:Vmax} and \ref{Fig:V} it follows that the
accelerating potential varies both in space and time. The spatial
variations occur between the two driftbands, over a range as large as
20\% of the polar cap radius; the temporal variations occur on
timescales much larger that pulsar rotation period. Still, both types
of variations are very small -- of order only several times
$10^{-4}\,\Vvac$.

If we next derive the nominal vacuum potential drop for \B\ per
eq.~(\ref{eq:Vvac}), using $\NSarg{R}{6}$=$1.0$, $B_{12}$=$1.4$ and
$P$=$1.85$, we find $\Vvac$=2.7$\times$$10^{12}$\,Volts. Thus, using
pulsar \B\ as a voltmeter indicates that the variations in its
acceleration potential are of the order of only $10^{9}$\,Volts.

\section{Discussion}
\label{Sect:Discussion}

Below we propose a qualitative model for drifting subpulses and
discuss potential future work.

\subsection{A model for drifting subpulses}

  The accelerating potential estimates we produce, only make the
  general assumption that subpulses are produced by emitting
  features that are stationary relative to the outflowing plasma in
  the polar cap region, and their drift is due to the rotation
  of this plasma, relative to the pulsar.  Here we wish to propose a
  more specific, but still qualitative and schematic, model for
  drifting subpulses that well fits the de-facto standard force-free
  model of pulsar magnetospheres, and that is consistent with the
  results presented in this paper.

\subsubsection{Timescales}

  First, we address the question why drift rates can change on
  timescales much larger than the pulsar's rotational period. In the
  frame of the force-free pulsar model, the potential drop in the
  pulsar polar cap depends on the current density flowing along a
  given magnetic field line and not only on the local physics of the
  polar cap
  \citep[e.g.][]{Levinson05,Timokhin2006:MNRAS1,Beloborodov2008,Timokhin2010::TDC_MNRAS_I}.
  Yet this current density depends on the {\em global} magnetospheric
  configuration
  \citep[e.g.][]{Timokhin2006,Timokhin2007:MNRAS2,timokhin::PSREQ_london/2006,BaiSpitkovsky2010a,Li2011,Kalapotharakos2011}.
  Changes in the accelerating potential in the polar cap should be,
  therefore, manifestations of changes in the global
  magnetosphere configuration. At some stages of the pulsar spin-down
  evolution, \citet{Timokhin_NULLING_2010} suggested, ulsar
  magnetosphere can change its configuration on timescales much larger
  than the pulsar period, and/or have meta-stable states.  In the
  standard model, the pulsar magnetosphere, the polar cap cascade
  zone, and the current sheet zone carrying most of the return
  current, are tightly coupled.  As these regions all have different
  time scales, the resulting system might constitute a highly
  non-linear dynamical system with varied behavior on a wide range of
  characteristic timescales \citep{Timokhin_NULLING_2010}.  Hence, the
  pulsar magnetosphere as a whole determines the properties of pulsar
  emission; such an idea in general was discussed before by
  \citet[][]{Wright2003} for example.

  So, when the global configuration of the magnetosphere changes, the
  current density distribution and the accelerating potential in the
  polar cap change with it, and hence does the plasma drift velocity
  in the open field-line zone.  Nonlinear dynamical systems can evolve
  on very long time scales, and so the pulsar magnetosphere might too
  change on time scales much larger than that of its composing
  parts. The pulsar under study here, \B\, is not unique in exhibiting
  changes on timescales many times larger than the rotation period:
  after some mode switches, pulsar B0943$+$10, for example, shows a
  gradual drift-speed change on a ∼4000-period time scale
  \citep{RankinSuleymanova2006}.  Conversely, if a magnetosphere 
  switches between meta-stable configurations, this results in mode
  changing and/or nulling. 
  Because these configurations have different current density
  distributions, we claim that with any mode change, the subpulse
  drift rate must change too. Several mode-changing
  pulsars are already known to exhibit such markedly different
  subpulse drift in different modes (e.g. PSR\;B0031$-$07,
  \citealt{smk05}; or PSR\;B1944+17, \citealt{kr10}).

\subsubsection{Magnitude of potential drop variations}

Given the accelerating potential in \B\, we next discuss how large the
inferred variations are, compared to its absolute value.  The maximum
achievable voltage in the polar cap is of the order of the potential
drop across it; the vacuum potential drop $\Vvac$ in our
terminology. At some point the onset of pair creation screens
  the electric field and limits the extent of the accelerating zone.
Especially in young pulsars the height of the acceleration zone is much smaller
  than the width of the polar cap, and the potential drop is much less
  than the corresponding $\Vvac$.

Now, recent self-consistent simulations of pair cascades in the polar
cap \citep{Timokhin2010::TDC_MNRAS_I,TimokhinArons::TDC/2011} have
shown that this pair-plasma generation is non-stationary: thus, that
each period of particle acceleration is followed by a quiet phase, in
which the accelerating electric field is screened and no pairs are
produced.  This holds for both the \citet{RudermanSutherland1975} and
space charge limited flow \citep{Arons1979} regimes. In these
simulations, the cascade behavior depends on the current density: that
density defines the duration of the active and quiet phases; it also
determines the maximum potential drop in the active phase.  Thus, these simulations
strongly suggest that traditional estimates for the potential drop
\citep[e.g.][]{RudermanSutherland1975,Hibschman/Arons:pair_multipl::2001}
are very inaccurate.

The source under study in this paper, \B, is a rather long-period
pulsar, with $P=1.85$~s, located close to the pulsar death line. If we
assume that the cessation of pair plasma production is responsible for
a pulsar's demise, then the potential drop in \B\ should be close its maximum
possible value: the vacuum drop. In Section~\ref{Sect:Modeling} we
found potential-drop variations with time of $<10^{-3}\Vvac$, i.e.\
very small compared to the value of the potential drop, even if the
actual potential drop would be an order of magnitude smaller than the
vacuum drop%
\footnote{Traditional estimates for the potential drop for \B,
  for dipolar and strongly non-dipolar (magnetic field-line curvature
  of $10^6$~cm) fields respectively, are $\sim$$\Vvac$ and $0.03\Vvac$
  for the space charge limited flow \citep[using expressions
  from][]{Hibschman/Arons:pair_multipl::2001}.  For the
  \citet{RudermanSutherland1975} model we get $0.6\Vvac$ and
  $0.04\Vvac$.}.

The magnetospheric configuration changes that cause such
one-in-a-thousand acceleration-drop fluctuations should be very small,
and are unlikely to influence the stability of the magnetosphere, or
be visible in profile or spindown-rate variations.  Overall, to us the
inferred small magnetospheric fluctuations seem quite plausible.

\subsubsection{Plasma and drifting subpulses}

Finally, let us now address the question of what could cause these
stationary features in the outflowing electron-positron plasma, that
manifest themselves as drifting subpulses.  The current most standard
explanation \citep{RudermanSutherland1975} assumes that plasma
generation in the pulsar polar cap occurs in ``spark columns'' that
are isolated by vacuum, and that these columns drift relative to the
NS. The plasma then flows only along certain magnetic field lines; and
thus the emitting regions are spatially localized.  In our view, such
a picture has serious difficulties. We will outline these below.

First, because of the magnetic field line curvature, any spark column
should quickly drift toward the magnetic pole, as pair-plasma
generation with each cascade iteration occurs closer to the magnetic
symmetry axis.  \citet{GilSendyk2000} have suggested that in a
specific symmetric configuration with a central spark, a polar cap
stably packed with sparks could exist.  More detailed simulations are
necessary to prove that this could indeed be the case even in such a
symmetric configuration.  Recent self-consistent simulation of pair
cascades \citep{Timokhin2010::TDC_MNRAS_I} have shown that the
characteristic time scale between discharges can be much larger than
previously assumed. This indicates that the electrodynamics of the
cascade zone, even for short period pulsars, resembles more strongly a
long tube with conducting walls, than a short cylinder with all
characteristic dimensions of order the polar-cap width $\PC{r}$ (as
assumed in \citealt{RudermanSutherland1975} models).  This would imply
that the coupling of the accelerating electric fields across the polar
is more complex than assumed before, and calling into question the existence of
distinct quasi-stable spark patterns such as those proposed in
\citet{GilSendyk2000}.

Second: we have thus far used eq.~(\ref{eq:v_drift}) on a polar cap
that is completely filled with plasma; the equation is, however, equally applicable to
plasma columns surrounded by vacuum. Outside of the plasma column the
electric field is unscreened, and the accelerating potential along the
magnetic field lines there is much higher than inside the column,
where the accelerating electric field is (partially) screened by the
pair plasma. Now the shape and variation of the potential drop
\emph{inside} the plasma column determine the plasma drift velocity.
Magnetospheric parameters like the magnetic field strength or the
current density should not change much over the width of the
column. The column is much smaller than the polar cap, and column
properties are set by the local physics such as the local plasma
density.  The magnetic field line in the plasma column which has the
maximum plasma density (the column ``center'') should then have the
smallest potential drop.  Thus, $V$ reaches its minimum
value here, and the plasma \emph{does not drift} there. The potential
drop rises from this minimum value toward all column edges.  Plasma on
opposite sides of the column ``center'', will drift in opposite
directions; a symmetric plasma column, for example, on average
does not drift at all, which is striking.  This invariably raises the
question of 
spark-column stability, as such differential rotation leads to
azimuthal smearing of the spark column. Of course, our arguments rely
on a rather qualitative picture for the plasma column, and only
accurate multidimensional simulations of pair cascades can verify
them, but they are nevertheless more rigorous than the original
\citet{RudermanSutherland1975} argumentation.
  
Based on these arguments we speculate that, in time-average sense, in
the active zone of the polar cap, pair productions happens uniformly,
without vacuum regions along some magnetic field lines.  The
potential-drop variation across the active zone is therefore rather
small and so the plasma drift is slow.  The distinct emitting features
in such plasma flow could be due to current filaments, somewhat
similar to ones observed in auroras.  If so, the generation of a
quasi-stationary system of such filaments depends on the global
magnetosphere structure, which could then also explain its
longevity.

Formation of a stable filament pattern may require a specific
magnetosphere configuration, thus happening in only some pulsars.  For
pulsars where stable filament structures cannot form (i.e.\ where
filaments form and disappear chaotically), there will be no clearly
visible drifting subpulses. In those pulsars however, steady plasma
rotation still occurs. And indeed, most of the pulsars studied by
\citet{wes06,Weltevrede2007} that do not show clear, stable drift
bands, still have periodic features in their power spectra -- in our
interpretation manifestations of the underlying {\em plasma rotation}.

\subsection{Future work}

Certain observations could verify or falsify our model.  

In this paper we relied on the geometrical model of \B\ proposed by
\citet{elg+05}.  Only with a more accurate geometrical model can
we place more robust constraints on the plasma rotation in the polar cap
on this pulsar. Polarization measurements can constrain this emission geometry by
measuring the polarization angle sweep.  Such observations would have
to be undertaken with sufficient collecting area, as only single-pulse
polarization measurements can identify and correct for the orthogonal
mode changes in \B\ \citep{lm88}.

Using their averaging technique, \citet{elg+05} followed the strong
drift tracks, and integrated all single pulses that have the same
drift track phase.  Those average profiles suggest that drift tracks
exist in between the two strongly emitting regions.  These tracks
sample co-latitudes between our two extremes $\xi_{1,2}$. Observations
with higher sensitivity than those presented here could measure the
pulse-to-pulse drift rate changes of those weak subpulses, and provide
acceleration potential estimate over the intermediate range of
colatitudes.

If observations at higher frequencies sample lower emission heights
\citep{cor78}, where the polar cap radius $\PC{r}$ is reduced, then
the observer sight line may start to cross $\PC{r}$. From that, the
extent of the polar cap and the values of $\xi_{1,2}$ could be
derived. In recent observations at 3094\,MHz \citep{sery11} the pulse
profile of \B\ still spans 360\,$^\circ$ however, so any evidence for
a decreasing pulse width would to be sought at even higher
frequencies.

After the report of the existence of the weak emission mode in \B\ by
\citet{elg+05}, no evidence for such a mode was found by
\citet{bgg08}, down to an upper limit reported to be more constraining
than the original detection in \citet{elg+05}. In addition to the
confirmation of the weak mode at 685 and 3094\,MHz in \citep{sery11},
we have here detected \B\ in weak mode in all datasets from Table
\ref{Table:Obs}; including in the data not originally used by
\citet{elg+05}, such as the dataset in the right panel of Figure
\ref{Fig:OnOff}.

In the model we suggest, the weak mode is a magnetospheric state that
is different from the strong mode, and should have a different plasma
drift rate. Thus, the detection of subpulse-drift in the weak mode would 
provide a stringent test of the general model we propose.

\section{Conclusions}
\label{Sect:Conclusions}

In this paper we pointed out that plasma rotation in the open magnetic
field-line zone depends not on the value but on the {\em variation} of
the accelerating potential across the polar cap.  If drifting
subpulses are caused by this plasma drift, then by measuring the drift
rate at two different colatitudes one can set limits on the
accelerating potential variation between these colatitudes.  We next
applied this technique to observations of aligned rotator \B, a
favorably oriented but otherwise average regular pulsar. Thus the
limits on the potential drop variation we obtained are representative
for $\sim1$~sec pulsars.

We found that the accelerating potential varies over the colatitude
range $\sim{}0.2\PC{\theta}$ by a factor of only several $10^{-4}$ of
the vacuum potential drop over the polar cap.  In \B, drift rates
change with time, suggesting that the accelerating potential along a
given magnetic field line changes with time as well.  The temporal
variations of the potential drop are again a few times $10^{-4}$ of the
vacuum potential over the polar cap, of the same order as the
potential difference across the colatitude range of the driftbands.

The smallness of these variations points to a remarkable stability of
the potential drop over these colatitudes, and provides useful
constraints on (future) self-consistent models of plasma
generation in pulsars.

~\\

\section*{Acknowledgments}
We thank Jonathan Arons for helpful discussions. JvL was supported by
the European Commission (grant FP7-PEOPLE-2007-4-3-IRG \#224838).  AT
was supported by an appointment to the NASA Postdoctoral program at
NASA/Goddard Space Flight center, administered by ORAU and also by NSF
grant AST-0507813; NASA grants NNG06GJI08G, NNX09AU05G; and DOE grant
DE-FC02-06ER41453.

\end{document}